\documentclass[12pt]{iopart}

\usepackage{graphicx,fixmath}
\begin{document}

\title[Centrality Dependence of Azimuthal Anisotropy of
Strange Hadrons]{Centrality Dependence of Azimuthal Anisotropy of
Strange Hadrons in 200\,GeV Au+Au Collisions}

\author{M Oldenburg for the STAR Collaboration}

\address{
Lawrence Berkeley National Laboratory, Nuclear Science Division \\1
Cyclotron Road, Berkeley, CA 94720, USA \footnote{current address:
European Organization for Nuclear Research, CERN, 1211 Geneva 23,
Switzerland} } \ead{Markus.Oldenburg@cern.ch}

\begin{abstract}
Measurements of azimuthal anisotropy for strange and multi-strange
hadrons are presented for the first time in their centrality
dependence. The high statistics results of $v_2(p_T)$ allow for a
more detailed comparison to hydrodynamical model calculations.
Number-of-constituent-quark scaling was tested for different
centrality classes separately. Higher order anisotropies like
$v_4(p_T)$ are measured for multi-strange hadrons. While we observe
agreement between measured data and models a deeper understanding
and refinement of the models seem to be necessary in order to fully
understand the details of the data.
\end{abstract}


\section{Introduction}
The initial stage of high-energy heavy-ion collisions can be probed
by measuring the azimuthal momentum anisotropies of the emitted
particles \cite{flow}. In non-central collisions the initial spatial
anisotropy of the reaction region is transformed into an collective
anisotropic motion, given that the constituents of the hot and dense
system interact with each other. Multi-strange hadrons give an
additional insight into the early phase of the reaction.
Measurements show \cite{blastwave} that these particles freeze out
with a lower mean collective velocity $\langle\beta_T\rangle$ and at
a higher temperature $T_{\rm{kin}}$ compared to protons, pions, and
kaons, suggesting that they decouple earlier from the strongly
interacting system. The measured significant azimuthal momentum
anisotropy in the final state must therefore be developed very
early, presumably during a partonic stage of the system. This effect
is called partonic collectivity.

To measure the final state momentum anisotropy the transverse
momentum $(p_T)$ distribution is expanded in terms of a Fourier
decomposition. The different Fourier coefficients $v_n = \langle
\cos(n\cdot\phi)\rangle$ are called `flow of order $n$'. The second
order coefficient $v_2$ is called `elliptic flow'.

The Au+Au data sample at $\sqrt{s_{\rm{NN}}}= 200\,\rm{GeV}$ used in
this study contains 13.3\,M minimum bias triggered events. This was
subdivided into three different centrality classes: 40--80\,\%
(peripheral) with 6.6\,M events, 10--40\,\% (mid-central) with
5.0\,M events, and 0--10\,\% (central). The number of events in the
central bin was enhanced to a total of 19\,M events by using our
large data set of central triggered events. The average event plane
resolutions for these three centrality regions were 66\,\%
(peripheral), 82\,\% (mid-central), and 69\,\% (central). The
average minimum bias event plane resolution was 76\,\%.

While a detailed study of the systematic errors is still under way,
initial estimates have been obtained for $K^0_S$ and
$\Lambda+\overline{\Lambda}$. The systematic errors are mainly
influenced by so-called non-flow effects: two particle correlations
not correlated with the event plane. Different analysis methods
(e.\,g.\ the cumulant method or analyses with Lee-Yang-zeros) offer
ways to reduce this non-flow contribution to the measured signal. By
comparing our results for the aforementioned particles obtained with
the `standard' method with the measurements from the different
methods it was established that at $p_T<4\,{\rm{GeV}}/c$ the
systematic uncertainty due to non-flow effects is of the order of
5\,\%, rising to 25--30\,\% at $4\,{\rm{GeV}}/c < p_T <
6\,{\rm{GeV}}/c$. Above $p_T = 5\,{\rm{GeV}}/c$ there is possibly a
large non-flow contribution. However, limited statistics prevents us
from being more quantitative. In view of these caveats only
statistical uncertainties are shown in this publication.

\section{Centrality dependence of $\mathbold{v_2(p_T)}$}
\begin{figure}[htb]
\includegraphics[width=1.05\linewidth]{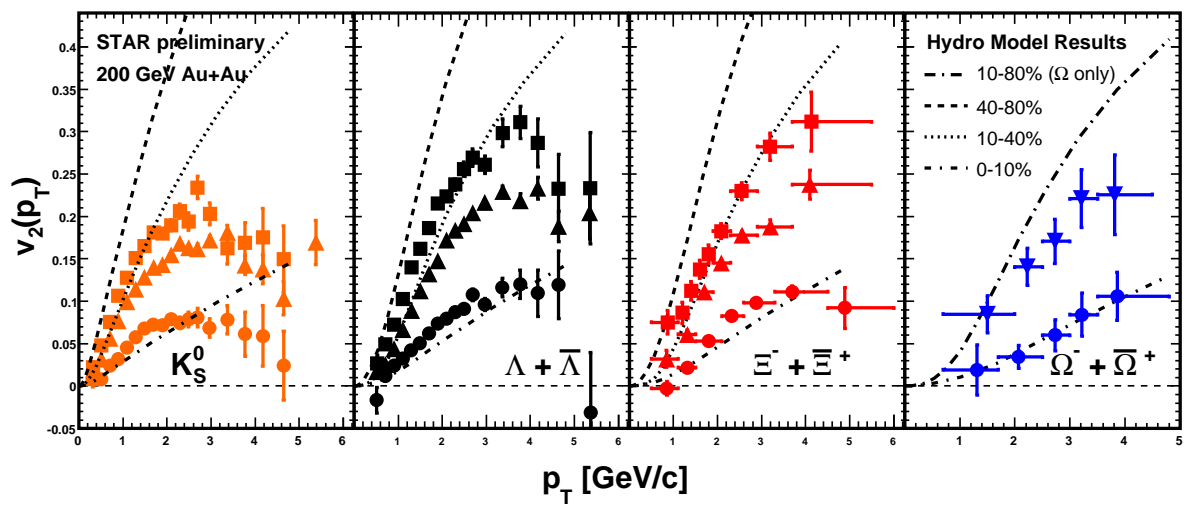}
\caption{\label{fig:1}Centrality dependence of $2^{\rm{nd}}$-order
azimuthal anisotropy. The panels show $v_2(p_T)$ for (from left to
right) $K^0_S$, $\Lambda+\overline{\Lambda}$, $\Xi+\overline{\Xi}$,
and $\Omega+\overline{\Omega}$. Different symbols represent
different centrality classes: 40--80\,\% (squares), 10--40\,\%
(triangles), 0--10\,\% (circles), and 10--80\,\% (upside down
triangles, for $\Omega+\overline{\Omega}$ only). Hydrodynamical
model calculations are also shown for the given particle species and
the same centrality regions.}
\end{figure}

Measurements of minimum bias $v_2(p_T)$ showed \cite{v4}
that in the low momentum region, where hydrodynamical effects
dominate, the strange and multi-strange particles show the same mass
ordering behavior as all the other particle species. In particular
this mass dependence can be described reasonably well with hydro
model calculations \cite{PasiPrivate}.

The available much higher than previous statistics allows for
measuring the centrality dependence of elliptic flow for strange and
multi-strange particles, see Fig.\,\ref{fig:1}. For $K^0_S$,
$\Lambda+\overline{\Lambda}$, and $\Xi+\overline{\Xi}$ three
centrality classes are shown. The lower yield of
$\Omega+\overline{\Omega}$ limited the measurement to two centrality
bins, only.

Even though the general trend of hydrodynamical model calculations
carried out for the same centrality regions is similar to the data,
we observe some significant deviations. Especially at low $p_T$,
where the biggest trust was put into these models, these differences
show that the overall agreement of the minimum bias measurements to
the model does not hold for this more detailed comparison. At higher
$p_T$ and for the peripheral bins the agreement breaks down
completely. It is interesting to note that for different particles
the agreement with the model is best for different centrality
classes. For example, while for $\Omega+\overline{\Omega}$ the most
central bin (0--10\,\%) shows the best agreement, the hydro
calculations fail to explain the results of all the other shown
particles at this centrality.

\section{Centrality dependence of number-of-constituent-quark scaling
of $\mathbold{v_2(p_T)}$}
\begin{figure}[htb]
\includegraphics[width=\linewidth]{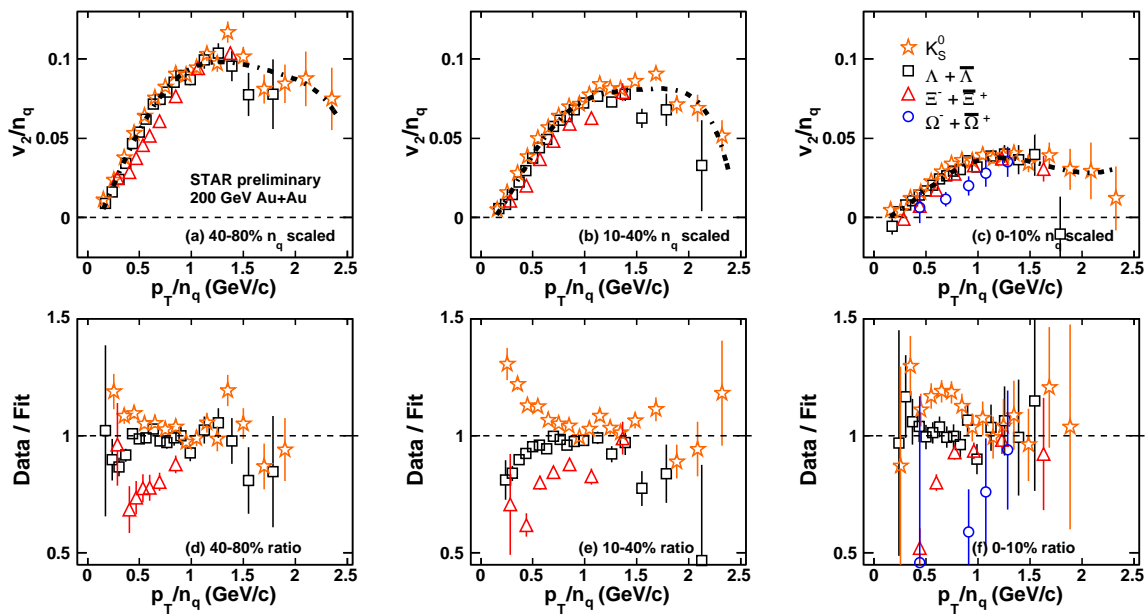}
\caption{\label{fig:2}Centrality dependence of $2^{\rm{nd}}$-order
azimuthal anisotropy, scaled by the number of constituent quarks for
$K^0_S$, $\Lambda+\overline{\Lambda}$, $\Xi+\overline{\Xi}$, and
$\Omega+\overline{\Omega}$. The different panels show different
centrality classes: a) and d) 40--80\,\%, b) and e) 10--40\,\%, c)
and f) 0--10\,\%. In the upper panels a common polynomial fit to all
data points is shown. For each centrality region the lower panels
show the ratio of the measured data to this common fit.}
\end{figure}

The scaling of $v_2(p_T)$ with the number of constituent quarks at
intermediate $p_T$ was established early on \cite{PaulOrSomething}.
Recently these measurements included multi-strange baryons as well,
even though only for minimum bias collisions. Our latest
measurements in Fig.\,\ref{fig:2}, upper row show the results for
$v_2(p_T)/n_{\rm{q}}$ vs.\ $p_T/n_{\rm{q}}$ for three different
centrality classes. Here $n_{\rm{q}}$ is the number of constituent
quarks of the given particle, i.\,e.\ $n_{\rm{q}} = 2$ for mesons
and $n_{\rm{q}} = 3$ for baryons. A common polynomial fit to all
data points is shown for each centrality bin.

The lower panels of Fig.\,\ref{fig:2} show the ratio of the measured
data points to the common fit for each centrality bin. While
especially for the most central collisions the statistics is not
sufficient the general trend is clearly visible: even in their
centrality dependence the $v_2(p_T)$ for different strangeness
containing particles seem to obey number-of-constituent-quark
scaling at intermediate $p_T$.

\section{Fourth order anisotropy $\mathbold{v_4(p_T)}$}
Higher statistics allows for measuring higher harmonics $v_n$, with
$n>2$, where the signal is usually smaller than $v_2$ and therefore
more difficult to distinguish from zero. The first measurement of
$v_4(p_T)$ for $\Xi+\overline\Xi$ shows strong, almost linear $p_T$
dependence. While ideal fluid dynamics predicts the ratio of $v_4$
to $v_2^2$ to be $1/2$ \cite{BorgKolb}, a simple comparison shows
that the ratio we observe is much better described by $v_4/v_2^2\sim
1.2$. This results is consistent with the earlier measurements
\cite{v4} for charged hadrons.

Even though it was argued that this deviation from ideal fluid
dynamics could be a hint of incomplete thermalization
\cite{Nicolas}, these results have to be taken with caution, because
the studies of the systematic errors and the contribution of
non-flow effects for the higher harmonics are still under way.

\section{Conclusions}
We have shown the centrality dependence of $v_2(p_T)$ for strange
and multi-strange hadrons. While the minimum bias measurement
follows the hydro model calculations, we observe a deviation from
the predictions for the different centrality classes. The high
statistics allowed for a more detailed study of
number-of-constituent-quark scaling, where at intermediate $p_T$ an
indication of scaling even for different centrality regions was
observed. Finally, the higher harmonic $v_4(p_T)$ was measured for
$\Xi+\overline\Xi$, showing similar scaling with $v_2^2$ as for the
other charged hadrons.


\section*{References}
\begin{thebibliography}{10}
\bibitem{flow} Poskanzer A M and Voloshin S A Phys.\ Rev.\ C {\bf 58} (1998) 1671. %
\bibitem{blastwave} Adams J {\it et al.} (STAR collaboration) Phys.\ Rev.\ Lett.\ {\bf 92} (2004) 112301;\\
    Adams J {\it et al.} (STAR collaboration) Phys.\ Rev.\ Lett.\ {\bf 92} (2004) 182301. %
\bibitem{v4} Adams J {\it et al.} (STAR collaboration) Phys.\ Rev.\ C {\bf 72} (2005) 014904. %
\bibitem{PasiPrivate} Huovinen P private communication (2005). %
\bibitem{PaulOrSomething} Adams J {\it et al.} (STAR collaboration) Phys.\ Rev.\ Lett.\ {\bf 92} (2004) 052302. %
\bibitem{BorgKolb} Borghini N and Ollitrault J Y arXiv:nucl-th/0506045; Kolb P F Phys.\ Rev.\ C {\bf 68} (2003) 031902. %
\bibitem{Nicolas} Bhalerao R S, Blaizot J P, Borghini N and Ollitrault J Y Phys.\ Lett.\ B {\bf 627} (2005) 49. %
\end{thebibliography}
\end{document}